# Effect of UV radiation on the structure of graphene oxide in water and its impact on cytotoxicity and As(III) adsorption

Waldo Roberto Gallegos-Pérez[a], Ana Cecilia Reynosa-Martínez[a], Claudia Soto-Ortiz[a], Mayra Angélica Álvarez-Lemus[b], Joaquín Barroso-Flores[c, d], Verónica García Montalvo[d], Eddie López-Honorato[a,*]

[a] Centro de Investigación y de Estudios Avanzados del IPN, Unidad Saltillo. Av. Industria Metalúrgica 1062, Parque Industrial. Ramos Arizpe, Coahuila, 25900, México.

[b] Universidad Juárez Autónoma de Tabasco, Av. Universidad s/n, Magisterial. Villahermosa, Tabasco, 86040, México.

[c] Centro Conjunto de Investigación en Química Sustentable UAEM-UNAM, Carretera Toluca-Atlacomulco Km 14.5, Unidad San Cayetano, Toluca, Estado de México, 50200 México.

[d] Instituto de Química. Universidad Nacional Autónoma de México, Circuito Exterior, Ciudad Universitaria, CD. MX., 04510, México.

**Abstract**

Graphene oxide (GO) is widely used in different applications, however once release into the environment it can change its structure and affect the transport of important contaminants such as arsenic. In this work we show that UV radiation, even in the range of 28-74 $\mu W/cm^2$ of irradiance up to 120 h of exposure, can induce important changes in the structure of graphene oxide, by eliminating -OH and C=O functional groups. This reduction affected the stability of graphene oxide in water by decreasing its zeta potential from 41 to 37 mV with the increase of the exposure time. Our results showed that after 24 h of UV exposure, As(III) adsorption capacity decreased from 5 mg/g to 4.7 mg/g, however after 48 h of irradiation the adsorption increased with time, reaching 5.1 mg/g at 120 h under 74 $\mu W/cm^2$ of irradiation. Computer modelling showed that even a degraded GO structure can have an interaction energy of 53 kcal/mol with $H_3AsO_3$. Furthermore, we observed that despite clear changes in surface composition and particle size, the reduction of graphene oxide maintained a high degree of cytotoxicity since cell viability decreased to 60% with a 50 µg/ml dose; except for the sample irradiated at 74 $\mu W/cm^2$ for five days, which showed 20% with the same concentration.

## 1. Introduction

Graphene oxide (GO) is a 2D nanomaterial with a network of carbon atoms with $sp^2$ and $sp^3$ hybridization and oxygen functional groups such as hydroxyl, carboxyl, and epoxy, which makes it very stable in water. GO has a wide range of applications in areas such as electronics, energy storage devices and as an adsorbent material for water purification among others.[1,2] Due to its continuous wide spread use, it is important to understand how graphene oxide behaves once released into the environment, particularly in water environments where it has been observed that UV irradiation can induce a reduction and breakage of this material. For example, Mohandoss et al. observed though X-ray photoelectron spectroscopy, the reduction in concentration of oxygenated functional groups, related to the detachment of hydroxyl (–OH), carboxylic acid (COOH) and epoxide (C-O-C) groups in the form of $CO_2$. Furthermore, Hou et al. observed similar behavior by monitoring the $CO_2$ formation by means of dissolved organic carbon (DOC). The release of $CO_2$ was related to a decrease in carbon content and an increase of defects that resulted in the fracture of the layers yielding a reduction in particle size.[3–5]

Furthermore, since it is known that the cytotoxicity of GO is affected by its degree of oxidation[6] and that GO by itself can work as adsorbent material via hydrogen bonds, and electrostatic and anion-π interactions,[7] it is of paramount importance to understand how these structural changes induced by UV irradiation might affect its cytotoxicity and possible transport of contaminants. For example, it has been observed that GO can increase the phytotoxicity of As(V) in wheat, but at the same time it can increase or decrease the toxicity of As(III), apparently depending on the as-produced GO structure.[5,7,8] However, it is important to mention that in these previous studies, pristine GO was used without considering the structural changes that GO undergoes once released into the environment. Since As is an element that affects millions of people worldwide, and whose chronical exposure is associated with skin, lung, liver and kidney cancer, among others,[10] it is import to study how the reduction and breakage of GO under UV irradiation affects the GO-As interaction.

In this work, we studied the effect of UV irradiance (28-74 µW/cm$^2$) up to 120 h of exposure, on the microstructure and composition of GO by combining experimental and modelling work. We observed that GO does undergo reduction by eliminating primarily -OH and COOH functional groups. However, despite clear changes in microstructure and composition, GO maintained in most cases its high degree of cytotoxicity and after 48 h of exposure showed a small increase on As adsorption. Our results are an example of the complex behavior GO once released into the environment.

## 2. Experimental

### 2.1. Graphene oxide synthesis

GO was prepared from the oxidation of graphite flakes (Sigma-Aldrich) using the improved Hummers method.[10,11] A solution of $H_2SO_4$ (95-98%, Jalmek)/$H_3PO_4$ (85.8%, J.T. Baker) with a 9:1 relation (180:20 ml) was first prepared. During constant agitation a mixture of graphite flakes/$KMnO_4$ (99%, Sigma Aldrich) in a relation 1:6 (1.5:9 g) was added. The resulting mixture was mixed under constant agitation during 15 minutes and then heat up to 50 °C for 24 h. After 24 h, the mixture was slowly cooled to 2 °C and 1.5 ml of $H_2O_2$ (30%, Jalmek) was added dropwise. The mixture was then taken to pH 1 by the addition of deionized water ($2\times10^{-6}$ ohm$^{-1}$cm$^{-1}$, Jalmek). The resulting solution was centrifuged (Premiere, XC-2450 series) at 3500 rpm for 15 minutes until the separation of the supernatant. The precipitate was then washed with deionized water, HCl (36.5-38%, Jalmek) and ethanol (99.5%, Jalmek) in consecutive sequence twice, and then coagulated with diethyl ether (99%, Jalmek). This last solvent was removed by heating the solution at 40 °C. Graphene oxide was then exfoliated in ethanol using an ultrasonic bath (Branson, 3800) for one hour. Finally, the GO was dried at 80°C for 12 h and ground in an agate mortar and sieve through a 100 mesh.

### 2.2. Photoreduction

The GO photoreduction was performed in 25 ml Flasks with water recirculation at 20°C. For this purpose, 0.0125 g of GO and 10 ml of deionized water were added per flask, at pH 7. These flasks were placed in a black box with 3 UV 7.2 watt lamps (Tecno Lite, F8T5BLB), with a wavelength of 368 nm. The intensity of the UV radiation was measured with a photodiode (OPHIR, PD-300 series) 15 cm of distance, giving values of irradiance of 28, 37, 54, and 74 µW/cm$^2$. The material was kept under constant magnetic stirring during 24 (UV1), 48 (UV2), 72 (UV3), 96 (UV4) and 120 hours (UV5) of exposure.

**2.3 Arsenic adsorption**

A standard solution of As(III) (1000 mg/L As(III) in 2% hydrochloric acid, Sigma-Aldrich) was used to prepare a solution with 25 mg/L of As, from which 10 ml were taken and placed in a flask with 0.0125 g of GO. Once the pH was adjusted to 7 the flasks were placed in the black box described previously, this time one of the flask was covered with aluminum foil to avoid UV irradiance on it. The adsorption experiment was carried at 24, 48, 72, 96 and 120 hours under constant magnetic stirring. Finally, all the solutions were centrifuged for 15 minutes at 3500 rpm and then filtered with a 0.45 and 0.2 µm poliethersulfone (PES) membrane (Whatman). Two drops of HNO$_3$ (66.3%, J.T. Beaker) were added to preserve the solution, according to the ASTM D2972-15 Norm for subsequent As quantification.

**2.4 Characterization**

The solid was analyzed with a Fourier Transform-Infrared Spectroscopy (FT-IR) on a PerkinElmer Frontier ATR-FTIR/NIR with a CPU32 Main software. Raman Spectroscopy was performed in a RENISHAW inVia Microscope using a laser excitation wavelength of 514 nm. X-Ray Photoelectron Spectroscopy was used on a PHI VersaProbe II with 2x10$^{-8}$ mTorr vacuum chamber, aluminum anode as X-ray monochromatic source and a radiation energy of 1486.6 eV. The analysis range was from 1400 to 0 eV. High resolution spectra of the C 1s signal was obtained for each of the samples. The high-resolution spectra were acquired with a step energy of 11.75 eV. The software used to do the deconvolution was CasaXPS (version 2.3.19PR1.0). Zeta Potential was measured in a Malvern Zetasizer Nano Z ZEN2600. As concentration of solutions was quantified by plasma atomic emission spectrometry (ICP, PERKIN ELMER optima 8300 model). Transmission electron microscopy analysis was performed in a FEI-Talos F200S with four in-column Super-X EDS detectors with a beam current of 300 pA and a collection time of 10 min.

## 2.5 Cell culture

Primary mononuclear cells were isolated using density gradient separation with the reagent Histopaque-1077 (Sigma-Aldrich, USA), according to the manufacturer′s instructions. All experiments were set up with mononuclear cells extracted from normal individuals (n=5). Cells were cultured with RPMI 1640 medium plus 10% of Fetal Bovine Serum, 10mM Penicillin Streptomycin, and 10mM of L-glutamine (Thermofisher Scientific, USA) in a 37ºC, 5% $CO_2$ humidified incubator. Cell viability was evaluated by the XTT (sodium 2,3,-bis(2-methoxy-4-nitro-5-sulfophenyl)-5-[(phenylamino)-carbonyl]-2H tetrazolium, Roche™) assay. Briefly, cells were seeded in a 96-well plate at a density of 5 x $10^3$ cells /well, and the compounds were incorporated into the medium at a concentration of 10, 30 and 50 µg/ml. The compounds employed for the treatments were freshly prepared in Dimethyl sulfoxide (DMSO); control cells were incubated only with DMSO. After 48 h, the medium was aspirated and cells were washed with PBS. Cells were cultured in a mix of 60 ml of DMEM/F12 medium without FBS plus 40ml of XTT-solution for 2 hours in a humidified incubator (at 37ºC, 5% $CO_2$). Production of formazan was quantified by using a microplate reader (Epoch, Bioteck), monitoring the difference in absorbance at 492 and 690 nm. All experiments were performed in triplicate. The effect of each treatment was expressed as a percentage of cell proliferation relative to untreated control cells, differences between control and treated groups are shown as a result of a two-way ANOVA statistics.

## 2.6 Computer Modelling

$H_3AsO_3$, $H_2AsO_3^-$ and $HAsO_3^{2-}$ were independently adsorbed over degraded graphene oxide and their geometries were optimized at the LC-ωPBE/6-31G(*d*,*p*) (Fig X) level of theory with the use of the Gaussian 09 suite of programs[13] The interaction energies were also calculated with the NBODel procedure with the NBO3.1[14] program as provided within the aforementioned suite, and the obtained values are collected in Table 3. The NBODel procedure deletes all orbital interactions between both species and the concomitant raise in energy is ascribed to the interaction between them.

## 3. Results and discussions

### 3.1. Changes in composition

Figure 1 shows an example the FTIR spectra of GO and GO after UV irradiation with different irradiance for 72 hours (UV3). In these spectra is possible to identify the signal of the stretching mode -OH at 3219 cm$^{-1}$, deformation mode of C=O at 1724 cm$^{-1}$, stretching mode of C=C at 1619 cm$^{-1}$, deformation mode of C−H at 1372 cm$^{-1}$ and flexion mode of C-O at 1036 cm$^{-1}$, particularly for the reference material, GO. As the UV irradiance increased from 28 to 74 µW/cm$^2$, the intensities of these bands changed, most of them decreased, giving the first sing of a structural transformation. For example, the most notorious changes are the disappearance of the –OH band and considerable reduction of the C=O band at 74 µW/cm$^2$. Conversely, the signals assigned to C=C and C-H bonds increased in relation to the other bands. Similar behavior is observed for GO irradiated for 24 and 120 hours (see Figure S1 in supporting information). Figure 1 shows that irradiance play an important role on GO reduction.

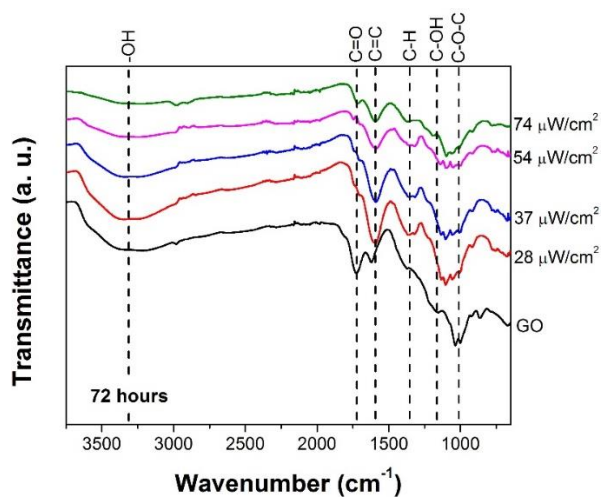

**Figure 1.** ATR-FTIR spectra of as-produced GO and irradiated GO UV3 (72 hours) at 28, 37, 54, 74 µW/cm$^2$.

Figure 2 shows the XPS spectra for carbon (1s) in the as-produced GO and GO irradiated with 37 µW/cm$^2$ during 72 and 120 hours. The as-produced GO showed the presence of C=C, C-C bonds (284.4 and 285.2 eV) attributed to the carbon structure with *sp$^2$* and *sp$^3$* hybridization, respectively. Additionally, hydroxyl (C-OH 286.4 eV), epoxide (C-O-C 287.1 eV), carbonyl (C=O 288 eV), and carboxylic acid (COOH 289.2 eV) functional groups,[11-14] with a C/O ratio of 1.35, were observed similarly to those previously identified by FTIR in Figure 1. The percentage of each type of bond is shown in Table 1. As the sample got irradiated for 72 h with 37 µW/m$^2$, the C=C and COOH bonds increased 20.6 and 1.2%, respectively. Conversely, the C-OH, C-O-C, O-C=O bonds decreased 16.1, 0.4 and 4.7%, respectively (Table 1), as also previously suggested by FTIR in Figure 1.

Similarly, after 120 h of irradiation with 37 µW/cm$^2$ the C-C bonds increased 22%, whereas the C-OH, C-O-C, O-C=O, COOH groups decreased 17.2, 0.6, 4.2, 1 %, respectively. It should be noted that compared to 72 hours of irradiation, the C-C bonds actually showed an increment of 22.5% (see Table 1), and the concentration of C=C bonds decrease from 32.6 to 12.8% from 72 to 120 h. These differences could be related to the bond energy needed to detach a functional group from the C structure.[18] For example, 360 kJ/mol are required to break a C-O bond, whereas 370 and 680 kJ/mol are needed for a C-C and C=C bond, respectively. Therefore, –OH, C-O-C, O-C=O and COOH are lost faster than the C-C and C=C bonds.[19] After 120 h of irradiation, due to the small concentration of oxygenated functional groups, carbon bonds appear to be altered.

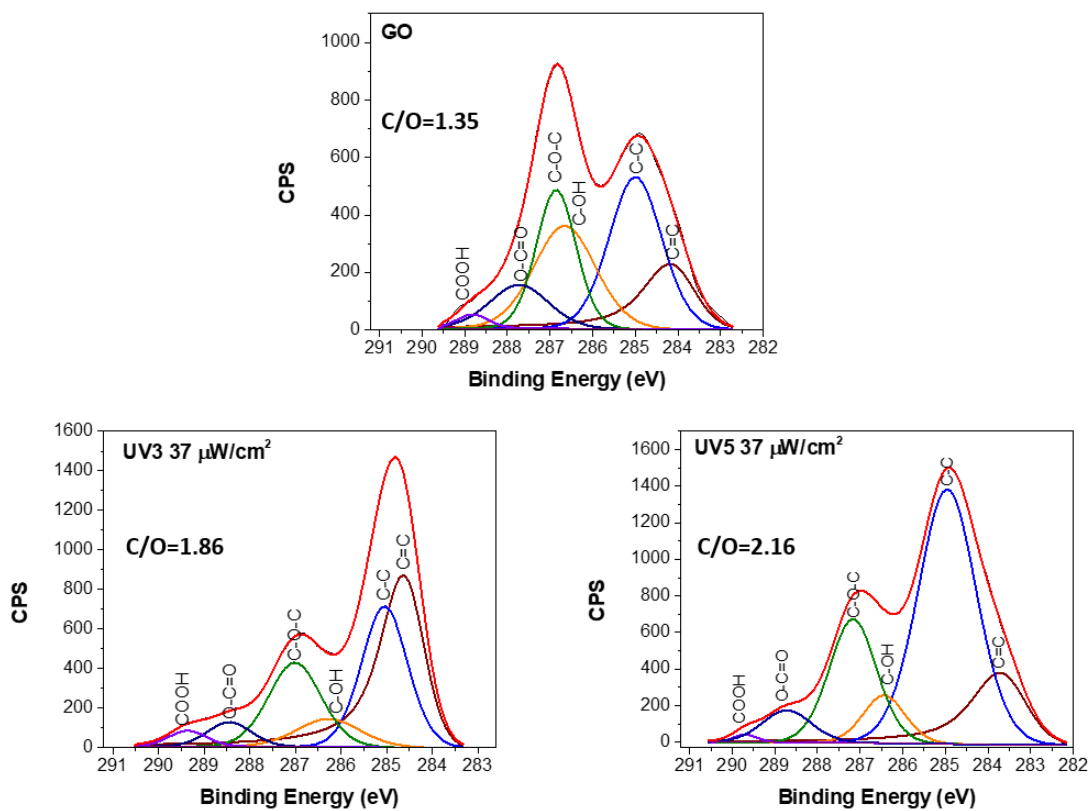

**Figure 2.** HR-XPS spectra for GO and irradiated GO, UV3 (72 hours) and UV5 (120 hours), at 37 µW/cm$^2$.

**Table 1.** Percentage of functional groups in GO and irradiated GO (UV3 and UV5) at 37 µW/cm$^2$ and 74 µW/cm$^2$.

| Type of bond | GO | 72 h (UV3), 37 µW/cm$^2$ | 72 h (UV3) 74 µW/cm$^2$ | 120 h (UV5) 37 µW/cm$^2$ | 120 h (UV5) 74 µW/cm$^2$ |
|---|---|---|---|---|---|
| | | % of bond | | | |
| C=C | 12.0 | 32.6 | 6.3 | 12.8 | 8.4 |
| C-C | 31.0 | 30.5 | 34.7 | 53.0 | 47.3 |
| C-OH | 24.5 | 8.3 | 30.1 | 7.2 | 12.5 |
| C-O-C | 21.2 | 20.8 | 23.9 | 20.8 | 28.0 |
| C=O | 9.8 | 5.0 | 4.0 | 5.6 | 3.2 |
| COOH | 1.7 | 2.8 | 0 | 0.7 | 0.6 |

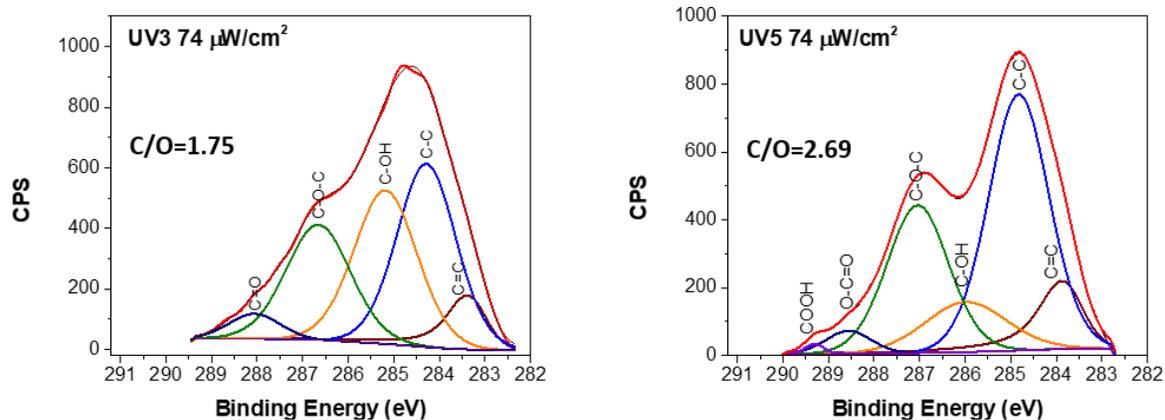

**Figure 3.** HR-XPS spectra for irradiated GO UV3 (72 hours) and UV5 (120 hours), at 74 µW/cm$^2$.

As can be seen in Fig. 3 and Table 1, as the UV radiation increase to 74 µW/cm$^2$ for 72 h, the C=C decreased 5.7% and the C-C, C-OH, C-O-C bonds increased 3.8, 6.6, and 2.7 % respectively, compared to the as-produced GO. However, the C=C bonds decrease 26.3% and the C-C bonds increased 4.3% as the irradiance increased from 37 to 74 µW/cm$^2$. Conversely, the C=O and COOH bonds decreased 5.7 and 1.7 %, respectively compared to pristine GO. It should be noted that C-O-C bonds had an increase of 3.1% as the UV irradiance increased during the first 72 h of irradiation. Similarly, as the irradiation time increased to 120 h with 74 µW/cm$^2$, the C=C bonds decreased 3.6 % and the C-C bonds increased 16.4 % compared to the as-produced GO, but with 4.4% and 5.7 % less than the value measured for 120 h with lower irradiance (37 µW/cm$^2$). Similar to the previous cases, the C-OH and C=O functional groups decreased 11.9, and 6.6 %, respectively, compared to pristine GO. Overall, it appears that longer periods of time under UV irradiation reduced the oxygen-bearing functional groups, as the C/O ratio changed from 1.4 of pristine GO up to 2.2 and 2.7 for 120 h of irradiation. However, as time increased up to 120 h, the C-O-C concentration seems to increase up to 7.2%, compared to the results obtained for 72 h of irradiation. The elimination of C-OH, C=O and C=O functional groups with an increase of C-C and C=C bonds, has also been previously observed by Hou et. al, even though the irradiance they used was 65 000 µW/cm$^2$, 10$^6$ times higher than the one used in this study.[17,18]

These changes in composition are expected to have a strong impact on the stability of this material in suspension. The measurement of the zeta potential (ζ) is a factor that describes this stability, which according to the literature values of ζ higher than 30 or -30 mV correspond to stable graphene oxide suspensions.[20] Figure 4 shows the effect of pH, irradiance and time on ζ. It is observed that at any given pH the as-produced GO is stable in water as for all the pH studies the zeta potential was below -30 mV. However, at pH 3 it is observed that the ζ of most of the irradiated GO crosses the stability line with values higher than -30 mV, while the as-produced GO remains stable. Similarly, it is observed that at pH values higher than 3, the stability of the irradiated materials increases with pH.

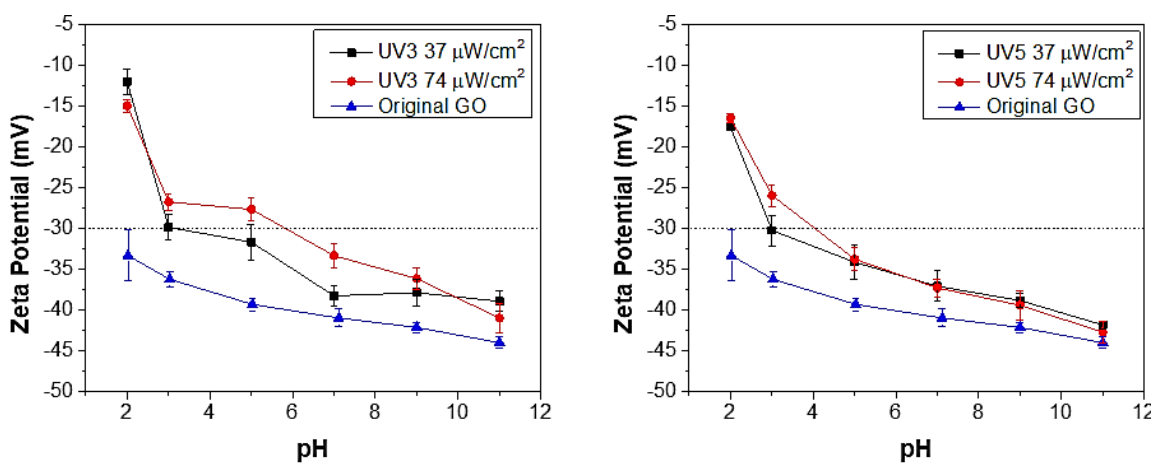

**Figure 4.** Zeta Potential of GO and GO irradiated 72 h (UV3) and 120 h (UV5) at 37, 74 µW/cm$^2$ with pH from 2 to 11.

By increasing the irradiance of GO from 37 to 74 µW/cm$^2$, a lower stability is achieved, particularly at pH 5. These changes in zeta potential agree with the results obtained by FTIR and XPS, suggesting the loss of functional groups, particularly C-OH and C=O, which confer its hydrophilic property to GO and allows it to remain in a stable suspension.[21] Furthermore, apparently, regardless of the clear changes in composition detected by FTIR and XPS, irradiated GO maintains certain degree of stability even after 120 h of irradiation.

Our results suggest that as the composition of the material progress, its stability in suspension is also affected, therefore, even though low irradiance in water might not mean a full transformation into reduced graphene oxide, it confers important structural changes that with time affect its composition and therefore its surface charge. Consequently, as time progresses and functional groups are loss, GO will become less and less hydrophilic, thus favoring the formation of agglomerates and promoting the material to resurface. As this happens a higher level of irradiance will be received thus accelerating this reduction and decomposition.

### 3.2 Microstructural changes

Raman spectroscopy is widely used for the characterization of graphene based materials. In Figure 4 it can be observed the Raman spectra of as-produced GO and irradiated GO at 36 and 74 µW/cm$^2$ during 72 and 120 h, all spectra show G band which is attributed to the $E_{2g}$ vibrational mode, corresponding to the relative movement of carbon atoms with $sp^2$ hybridization in both rings and chains. [21–23] This band is present too in graphite in ~1580 cm$^{-1}$. In GO besides this bands there is D band too, associated with the vibrational mode $A_{1g}$ generated by the disorder in carbon layer in plane and in edges.[21,24] Some authors use the ratio between these bands ($I_D/I_G$) to estimate the defects degree in the graphene network, we used this ratio as an indicator of oxidation degree since the more oxidized GO the more defects will have.[25,26] Figure 5 shows the Raman spectrum of the as-produced GO, with a $I_D/I_G$ ratio of 1.18 ± 0.08. After irradiation for 72 hours (Figure 5), the $I_D/I_G$ ratio had a slightly change of 1.14 ± 0.72 and 1.20 ± 0.11 for 37 and 74 µW/cm$^2$ of irradiance, respectively. Furthermore, Figure 5 shows the Raman spectra of GO irradiated for 120 h. After the longest time studied under irradiation, the $I_D/I_G$ ratios changed to 1.13 ± 0.06 and 1.17 ± 0.05 for 37 and 74 µW/cm$^2$, respectively.

Since the intensity of the D band is associate with the presence of defects (edge or in-plane),[28] an increase of the $I_D/I_G$ would represent the formation of a more disorder structure. Conversely, a reduction of the $I_D/I_G$ intensity ratio is generally associate with an increase of graphitization (higher areas with well-structured hexagonal carbon graphene structure). Although our results show a slight variation towards the formation of a more graphitic structure, the intensity ratio $I_D/I_G$ did not show clear evidence of structural changes of graphene oxide as were clearly observed by FTIR and XPS. This could be related to a competing behavior between the restructuring of the $sp^2$ as the oxygenated functional groups are removed from the structure (would reduce the D band intensity) but also, the continuous fracture of the material and the formation of $sp^3$ structures at the edge of the material (would increase the D band intensity).[3]

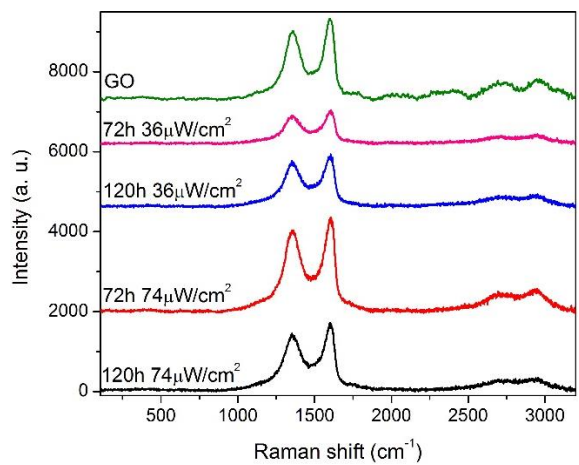

**Figure 5.** Raman spectrum of GO irradiated for 72 h (UV3) and 120 h (UV5) at 37 µW/cm$^2$ and 74µW/cm$^2$.

The structural changes suggested by Raman spectroscopy were also visible by TEM (Figure 6). TEM micrographs of as-produced GO (Figure 6a) showed that the original material had single layers of approximately 25 µm in length with soft edges and distinctive winkles, particularly at the edge. This tortuosity is distinctive to GO due to the double carbon vacancy ($C_2$) defects which generates non-regular rings commonly 5-8-5 member rings, but also Stone-Wales defects formed by a pair of adjacent 5-7 member rings increasing tension in carbon structure,[27,28] besides the formation of carbons with *sp$^3$* hybridization results in the tortuosity of the carbon structure. Overall, most of the structure of as-produced GO is amorphous as can be seen in Fig. 6b.

However, as the sample got irradiated with 74 µW/cm$^2$ for 120 h, GO changed considerably as it can be seen in Figure 7a, where the distinctive single layer structure of GO was no longer observed. Instead, a large number of irregular particles with rough edges between 20 nm to 8 µm in length were detected. At higher magnification, it was possible to identify that some of the particles were in fact agglomerates of smaller particles of around 100 nm (Figure 7b) and that the irradiated graphene oxide layers were severely defective at the edges and on the plane as a results of the decomposition process induced by UV irradiation.[31] These new structures showed clear evidence of higher crystallinity as can be observed in the dark field images (Figure 7c) where the bright sections correspond to crystalline structures with similar orientation.[32] This increase in crystallinity was also evident in higher resolution TEM where graphene domains were now visible, albeit still in a random arrangement (Figure 7d and 7e). These results suggest that graphene oxide suffered partial reduction but a considerable degradation and formation of smaller particles with sharp edges.[33]

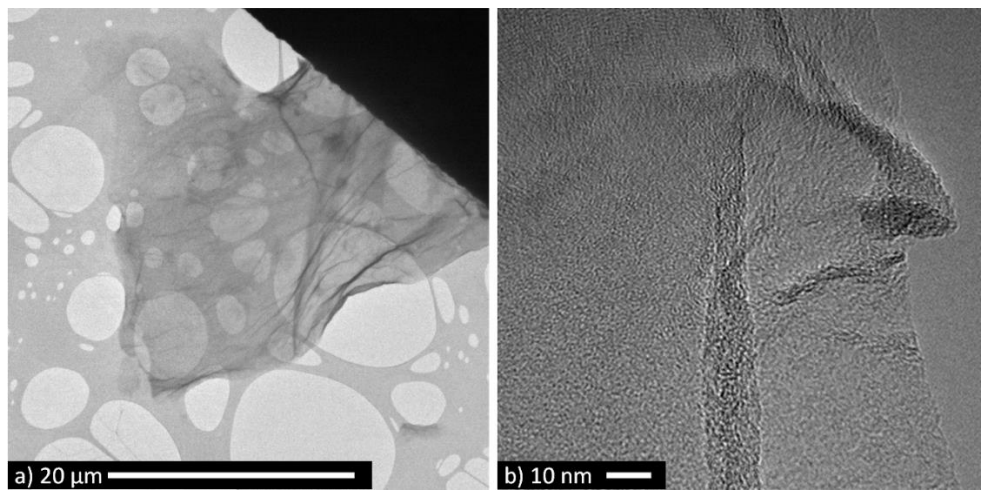

**Figure 6.** TEM micrographs of as-produced GO.

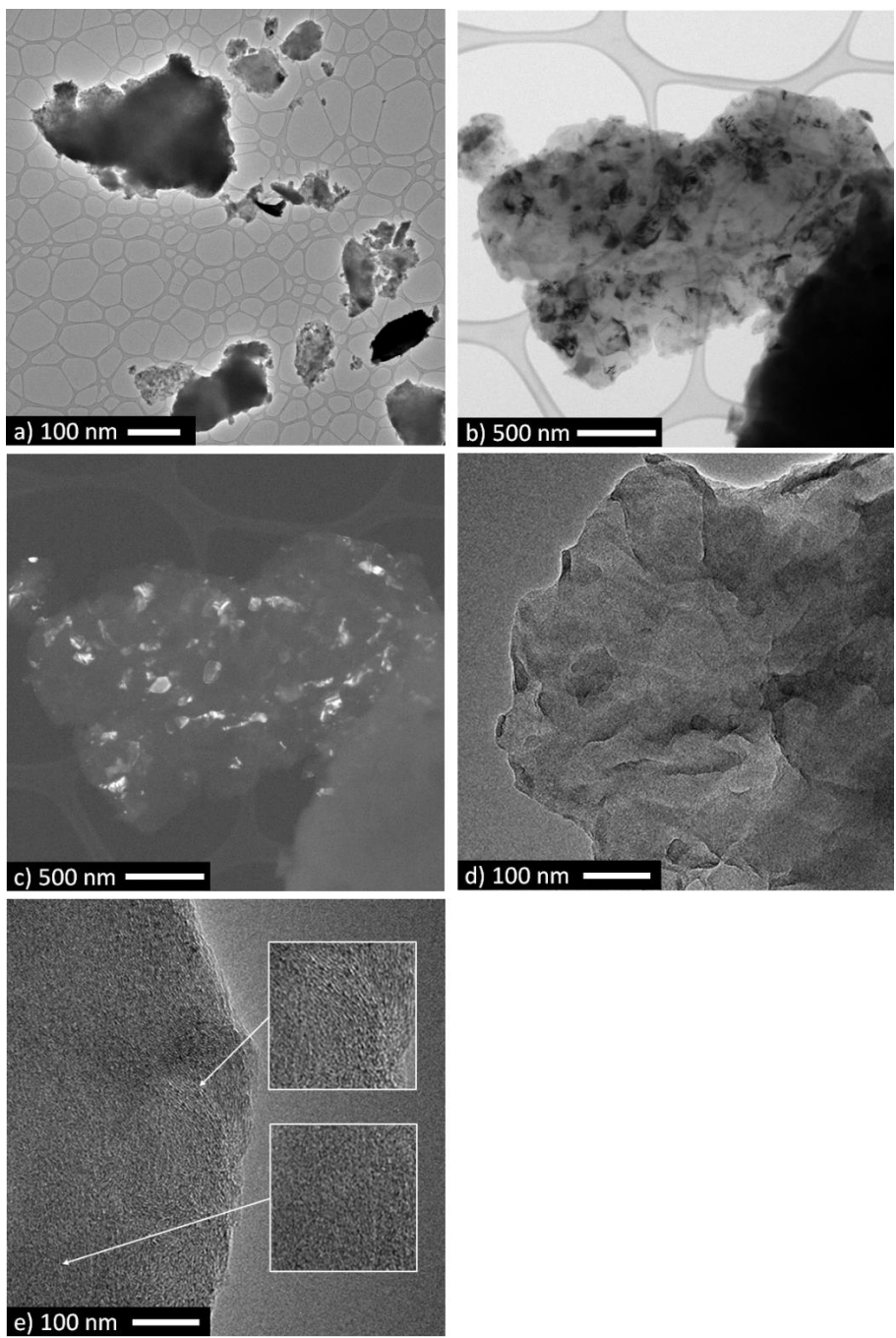

**Figure 7.** TEM micrographs of irradiated GO with 74μW/cm$^2$ 120 h. a) Low magnification general overview of GO, b) and c) bright and dark field image showing the presence of crystalline structures within GO agglomerates. d) and e) High magnification micrograph showing the presence of individual particles and ordered structures within GO particles.

### 3.3 As(III) adsorption tests

Computer models suggest that GO interacts with As(III) through hydrogen bonds.[34] Therefore, any change in the structure and composition is expected to have an impact on As(III) adsorption and transport. Originally, the as-produced GO had an adsorption capacity of 5.02 mg/g As(III) at pH 7. However, after 24 h of irradiation, the As(III) adsorption dropped to 4.7 and 4.8 mg/g for 74 and 37 µW/cm$^2$ of irradiance, respectively (Figure 8). Overall, higher irradiance resulted in lower adsorption capacities. Furthermore, our results show that As adsorption of irradiated GO was almost the same until 72 h, time after which its adsorption appeared to increased once again. Surprisingly, both GO with 37 and 72 µW/cm$^2$ of irradiance after 72 h reached 4.9 mg/g of As(III) adsorption. It was after 120 h of irradiance that adsorption capacity reached values of 4.9 and 5.1 mg/g for 37 and 74 µW/cm$^2$, respectively. It appears that time plays an important role on As(III) adsorption under UV irradiation.

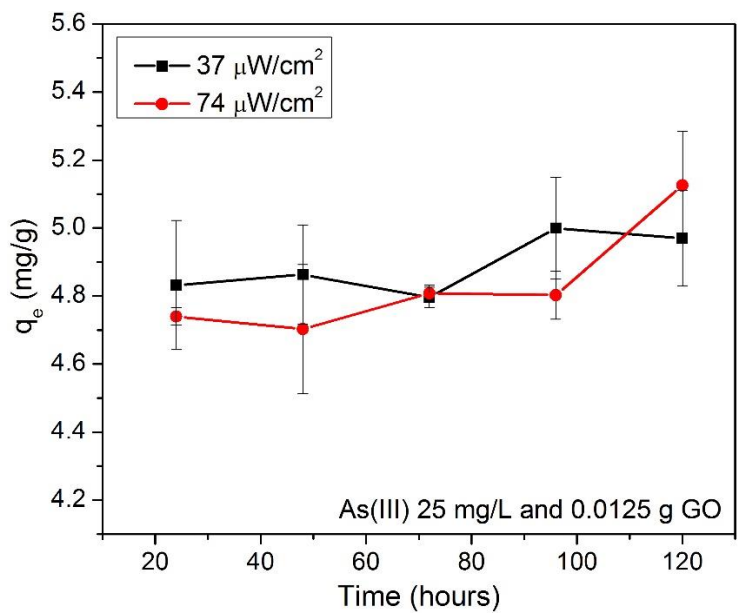

**Figure 8.** Adsorption capacity of irradiated GO to As (III) in relation to the time of exposure to UV radiation. Initial GO adsorption capacity was 5.02 mg/g As(III).

It has been previously been suggested that UV irradiation can result in the degradation of GO by induced photolysis through the formation of holes in the basal plane and the concomitant formation of carbon atoms with *sp$^3$* hybridization as functional groups are reduced by UV irradiation.[32,33] This reduction process could be done through the oxidation of pristine $sp^2$ carbon regions with hydroxyl groups by its reaction with highly reactive hydroxyl radicals and the formation of epoxy and carbonyl groups as an intermediate step in the reduction of the hydroxyl functional groups.[31] Furthermore, it has been observed that graphene can also be oxidized when irradiated and form primarily new C-O-C and C-OH functional groups.[37] This reduction route could explain the changes on C=O and -OH functional groups detected by XPS. Where C-C bonds concentration always increases with irradiance and time, but the concentration of -OH and -CO, change considerably depending on time and irradiance. For example, at 72 h of irradiation with 74 µW/cm$^2$ the concentration of -OH functional groups increased from 24 to 31%, however as time increased to 120 h, its concentration dropped to 12%. It appears that at short times and low irradiance in the range of 37 µW/cm$^2$, GO is reduced.[38] However, prolonged exposure or higher irradiance, allows an increase on oxygen content in GO as an intermediate step on the decomposition and reduction of GO, probably due to the formation of a higher concentration of hydroxyl radicals.[31] This can be seen in the increase of C-OH bonds detected by XPS at 72 h but the final reduction on C-OH concentration at 120 h. It is important to mentioned that in our experiments it is apparent that C-O-C concentration remained almost the same with even a slight increase on concentration as irradiance and time increased to 120 h and 72 µW/cm$^2$ (from 21% for GO to 28%). This could be related to epoxy ring opening/closing reactions induced by OH groups generated by UV irradiation.[31]

These variations in functional groups could explain the variations seen in adsorption capacity. Table II show the interaction energies for graphene ($sp^2$ carbon), epoxy and hydroxyl functional groups with the different As(III) species calculated for pristine GO. Graphene has a negligible interaction with arsenic ($H_3AsO_3$ since we worked at pH 7). The reduction on As(III) capacity of UV irradiated GO during the first 72 h could be related to the reduction of -OH functional groups, which have the strongest interaction with As. However, as irradiation and time increase, the remaining GO structure, in particular the constant presence of epoxy groups and the partial re-oxidations of GO debris, and even the newly form $sp^3$ structure increase once again the interaction with As. This can be seen in Figure 9 and Table 3 where a simulated degraded GO with a hole in its structure and the presence of $sp^3$ hybridization, reached interaction energies of 53 kcal/mol for the neutral $H_3AsO_3$ spicy. This suggest that even during the disordering of the structure GO-As interaction can increase compared to pristine graphene, but lower than the original GO structure.

**Table 2.** Interaction Energies [kcal/mol] calculated at the LC-ωPBE/6-31G(*d,p*) level of theory.

| Functional group | As (III) | | |
|---|---|---|---|
| | $H_3AsO_3$ | $H_2AsO_3^-$ | $HAsO_3^{2-}$ |
| Graphene | 3.90 | 254.54 | 383.35 |
| Epoxide | 360.57 | 329.98 | 0.72 |
| Hydroxyl | 378.50 | 383.35 | 378.98 |

**Table 3.** NBODel Interaction Energies [kcal/mol] of degraded GO and As(III)

| As(III) | Energy (kcal/mol) |
|---|---|
| $H_3AsO_3$ | 53.50 |
| $H_2AsO_3^-$ | 141.57 |
| $HAsO_3^{2-}$ | 17.53 |

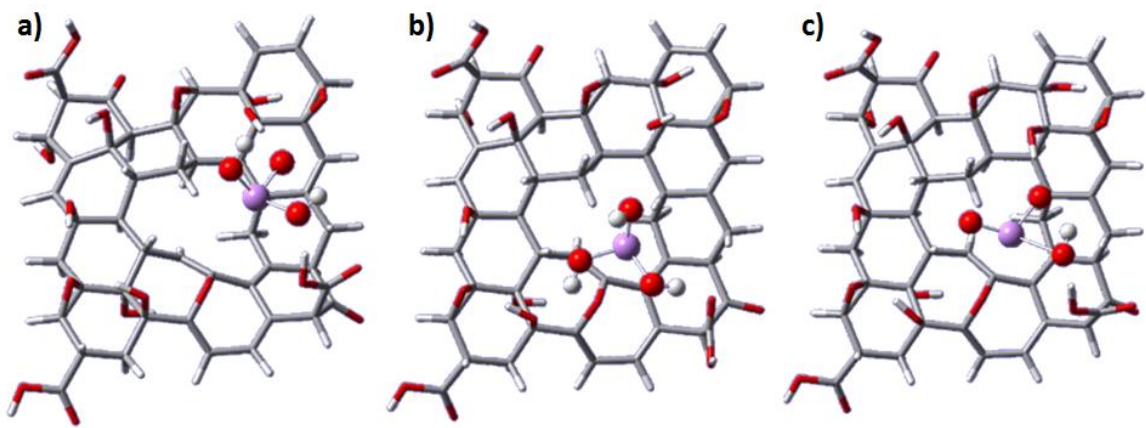

**Figure 9.** Optimized geometries for degraded graphene oxide complexes with $H_3AsO_3$ (*a*), $H_2AsO_3^-$ (*b*) and $HAsO_3^{2-}$ (*c*).

Our results show that the complex reduction and decomposition process of GO under UV-irradiation not necessarily means a lower capacity of transport of contaminants. As GO change in composition and structure, other heavy metal ion or pollutants might even have higher adsorption capacity since electrically charge molecules have a stronger interaction with also the carbon structure of GO as can be seen in Table 2 for $H_2AsO_3^-$ and $HAsO_3^{2-}$, As species that are present only at reducing conditions.

## 3.4 Cytotoxicity

As a result of UV irradiation the structure of GO underwent several modifications and is expected that its behavior in nature, specially its hazardness, change too. Figure 10 shows the dose-dependent effect of as-produced and irradiated GO on cell viability, where monocytes were used as test cells. At 10 μg/ml of non-irradiated GO the cell viability decreased by 20%, compared with the control test, whereas all irradiated GO materials decreased by ~10%. This difference in cytotoxicity reduced as concentration increased up to 50 μg/ml, where as-produced and irradiated GO induced almost the same effect on cell viability (60%). Nevertheless, only the samples irradiated for 120 h with 74 μW/cm$^2$ exhibited the lowest cytotoxic effect at the 30 and 50 μg/ml doses, since viability of the cells remained almost the same (80%) when compared to control. Overall, GO reduction and fracture does not necessarily reduce GO cytotoxicity, despite the huge differences in composition and particle size.

The cytotoxicity of GO has been observed to be strongly depended on particle size and oxygen content, as the oxygenated functional groups facilitate the interaction between the GO surface and the cell membrane.[39,40] Therefore, a lower concentration of oxygenated functional groups should have resulted in a lower cytotoxicity, however, a reduction of particle size and an increase of defects at the edge of these new GO debris (Fig. 7) could be responsible for the high cytotoxicity showed. Similar behavior has also been observed before for hydrazine reduced GO, which induced higher cytotoxicity than GO[6], presumably due to the existence of sharp edges. Furthermore, the reduction of oxygenated functional groups is generally concentrated along the plane in GO, whereas the formation of new oxygenated functional groups could happen at these new bonds at the edge. Figure 11 shows that despite the variations in concentrations detected by XPS, EDS at high resolution TEM show that the edge of GO debris still contain a relatively uniform distribution of oxygen. With the current results it is still unknown why the sample UV5/74 had lower cytotoxicity than the other samples, however further work is currently underway to elucidate the origin of this cytotoxicity as GO is degraded by UV-irradiation.

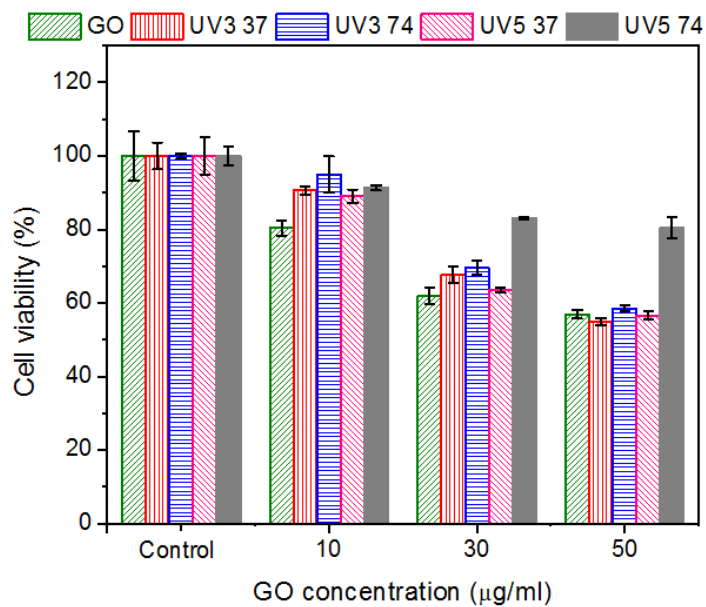

**Figure 10.** Cytotoxicity of as-produced GO and irradiated GO at 37 and 74 µW/cm$^2$ during 72 h (UV3) and 120 h (UV5).

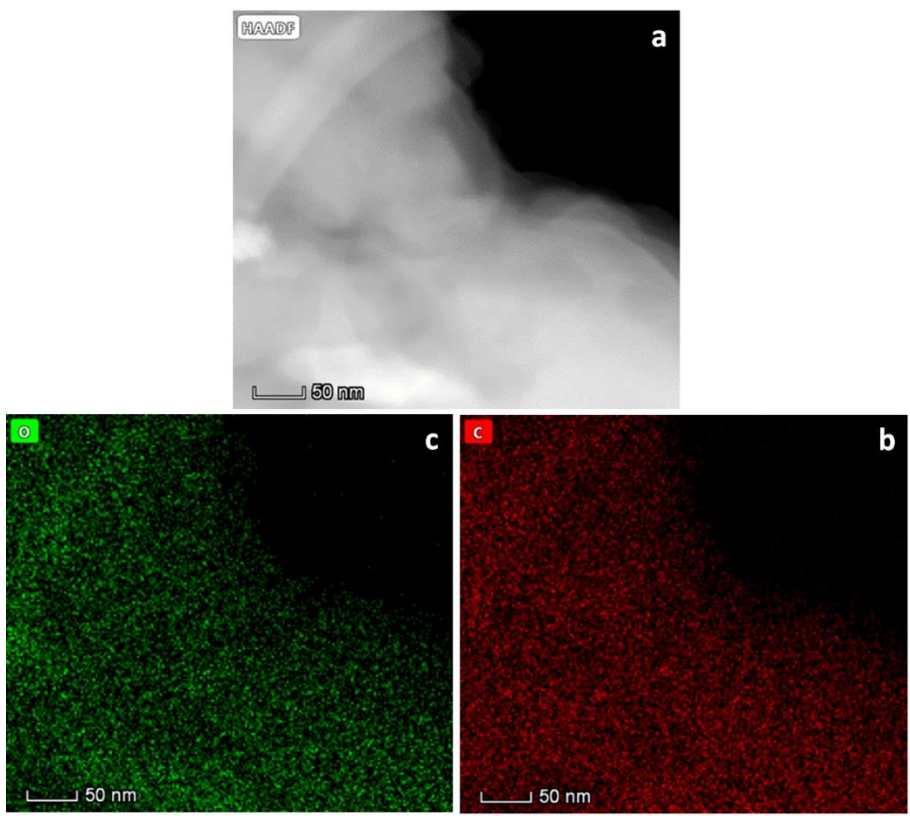

**Figure 11.** EDS elemental maps of GO after 120 h of 74 µW/cm$^2$. a)HAADF micrograph of the edge of a GO debris; b) carbon elemental map; c) oxygen elemental map.

**Conclusions**

Our results suggest that graphene oxide suffers of a partial reduction and decomposition under UV irradiation and that this process is strongly dependent on exposure time and irradiance. During short time and low irradiance (72h at 37 µW/cm$^2$) we observed by XPS and FTIR that GO is primarily reduced, losing primarily -OH and C=O functional groups. However, as time or irradiance increase, a process or re-oxidation appears to occurred as confirmed by XPS measurements who showed an increase of C-O-C functional groups. Furthermore, these variations in surface composition were also reflected on the capability of GO to adsorb As, particularly As(III). During the first 24 h of irradiation, GO reduce its adsorption capacity by 6.4%, however after 72 h of irradiance, it started to regain its adsorption capacity reaching values of 4.9 mg/g after 120 h of irradiation at 74 µW/cm2. TEM, XPS and computer models suggest that the re-oxidation of GO as well as the presence of C atoms with sp$^3$ hybridization can increase the GO-As(III) interaction, even for the neutral species of H$_3$AsO$_3$, which reached interaction energies of 53 kcal/mol, a higher energy compared to the sp$^2$ structure found in graphene which only had interaction energies of 3.9 kcal/mol. These changes in composition and crystallinity also were reflected on the cytotoxicity of the material. The cytotoxicity of GO decreased after 5 day of irradiation at 74 µW/cm$^2$, although the higher concentration was used, it can be related with the material reduction and its severe microstructural change observed in HR-TEM and the detachment of its oxygenated functional groups as indicated in XPS analysis.

**Conflict of interest**

There are not conflicts to declare.

**Acknowledgments**

This material is based upon work supported by a grant from the Consejo Nacional de Ciencia y Tecnología (CONACYT, project number: 247080, Problemas Nacionales). The authors would like to acknowledge CONACYT for the PhD, MSc and BSc grants awarded to A.C. Reynosa, W.R. Gallegos Perez and C. Soto-Ortiz, respectively.

**References**

1.  Pendolino F, Armata N. Graphene Oxide in Environmental Remediation Process [Internet]. 1 st. Kacprzyk J, Polish Academy of Science, Systems Research Instutute, editors. Warsaw, Poland: Springer International Publishing AG; 2017. 1-56 p. Available from: http://link.springer.com/10.1007/978-3-319-60429-9

2.  Zhu Y, Murali S, Cai W, Li X, Suk JW, Potts JR, et al. Graphene and graphene oxide: Synthesis, properties, and applications. Adv Mater. 2010;22(35):3906–24.

3.  Mohandoss M, Gupta S Sen, Nelleri A, Pradeep T, Maliyekkal SM. Solar mediated reduction of graphene oxide. RSC Adv. 2017;7(2):957–63.

4.  Hou WC, Chowdhury I, Goodwin DG, Henderson WM, Fairbrother DH, Bouchard D, et al. Photochemical transformation of graphene oxide in sunlight. Environ Sci Technol. 2015;49(6):3435–43.

5.  Li H, Bubeck C. Photoreduction processes of graphene oxide and related applications. Macromol Res. 2013;21(3):290–7.

6.  Liu S, Zeng TH, Hofmann M, Burcombe E, Wei J, Jiang R, et al. Antibacterial

Activity of Graphite , Graphite Oxide , Graphene Oxide , and Reduced Graphene Oxide : Membrane and Oxidative Stress. ASC Nano. 2011;5(9):6971–80.

7. Chen X, Chen B. Direct Observation, Molecular Structure, and Location of Oxidation Debris on Graphene Oxide Nanosheets. Environ Sci Technol. 2016;50(16):8568–77.

8. Hu X, Kang J, Lu K, Zhou R, Mu L, Zhou Q. Graphene oxide amplifies the phytotoxicity of arsenic in wheat. Sci Rep. 2014;4:1–10.

9. Dai H, Yun L, Wang J, Nie Y, Sun Y, Wang M. Graphene oxide antagonizes the toxic response to arsenic via activation of protective autophagy and suppression of the arsenic-binding protein LEC-1 in Caenorhabditis elegans. Environ Sci Nano. 2018;5(7):1711–28.

10. Choong TSY, Chuah TG, Robiah Y, Gregory Koay FL, Azni I. Arsenic toxicity, health hazards and removal techniques from water: an overview. Desalination. 2007;217(1–3):139–66.

11. Marcano DC, Kosynkin D V., Berlin JM, Sinitskii A, Sun Z, Slesarev A, et al. Improved synthesis of graphene oxide. ACS Nano. 2010;4(8):4806–14.

12. Ganguly A, Sharma S, Papakonstantinou P, Hamilton J. Probing the Thermal Deoxygenation of Graphene Oxide using High Resolution In Situ X Ray based Spectroscopies. J Phys Chem C. 2011;115(34):17009–19.

13. Frisch ML. Gaussian 09, Revision A.02. Wallingford CT: Gaussian, Inc; 2009.

14. Weinhold F, Glendening ED. NBO3.1. Wisconsin; 1996.

15. Shim SH, Kim KT, Lee JU, Jo WH. Facile method to functionalize graphene oxide and its application to poly(ethylene terephthalate)/graphene composite. ACS Appl Mater Interfaces. 2012;4(8):4184–91.

16. Mattevi C, Eda G, Agnoli S, Miller S, Mkhoyan KA, Celik O, et al. Evolution of electrical, chemical, and structural properties of transparent and conducting chemically derived graphene thin films. Adv Funct Mater. 2009;19(16):2577–83.

17. Stobinski L, Lesiak B, Malolepszy A, Mazurkiewicz M, Mierzwa B, Zemek J, et al. Graphene oxide and reduced graphene oxide studied by the XRD, TEM and electron spectroscopy methods. J Electron Spectros Relat Phenomena [Internet]. 2014;195:145–54. Available from: http://dx.doi.org/10.1016/j.elspec.2014.07.003

18. Mulyana Y, Uenuma M, Ishikawa Y, Uraoka Y. Reversible Oxidation of Graphene Through Ultraviolet/ozone Treatment and Its Non- thermal Reduction Through Ultraviolet Irradiation. J Phys Chem C. 2014;118:27372–81.

19. Glockler G. Carbon–Oxygen Bond Energies and Bond Distances. J Phys Chem. 1958;62(9):1049–54.

20. Hunter RJ. Zeta potential in colloid science: principles and applications. Vol. 2. Academic press; 2013.

21. Dai J, Wang G, Ma L, Wu C. Study on the surface energies and dispersibility of graphene oxide and its derivatives. J Mater Sci. 2015;50(11):3895–907.

22. Sethuraman VA, Hardwick LJ, Srinivasan V, Kostecki R. Surface structural disordering in graphite upon lithium intercalation/deintercalation. J Power Sources. 2010;195(11):3655–60.

23. Lin TY, Chen DH. One-step green synthesis of arginine-capped iron oxide/reduced graphene oxide nanocomposite and its use for acid dye removal. RSC Adv. 2014;4(56):29357–64.

24. Vollebregt S, Ishihara R, Tichelaar FD, Hou Y, Beenakker CIM. Influence of the growth temperature on the first and second-order Raman band ratios and widths of carbon nanotubes and fibers. Carbon N Y. 2012;50(10):3542–54.

25. Claramunt S, Varea A, López-Díaz D, Velázquez MM, Cornet A, Cirera A. The importance of interbands on the interpretation of the raman spectrum of graphene oxide. J Phys Chem C. 2015;119(18):10123–9.

26. Lõpez-Díaz D, Velázquez MM, De La Torre SB, Pérez-Pisonero A, Trujillano R, Fierro JLG, et al. The role of oxidative debris on graphene oxide films. ChemPhysChem. 2013;14(17):4002–9.

27. Gong Y, Qin C, He W, Qiao Z, Zhang G, Chen R, et al. Solar light assisted green synthesis of photoreduced graphene oxide for the highefficiency adsorption of anionic dyes. RSC Adv. 2017;7(4):53362–72.

28. Su Y, Du J, Sun D, Liu C, Cheng H. Reduced graphene oxide with a highly restored π-conjugated structure for inkjet printing and its use in all-carbon transistors. Nano Res. 2013;6(11):842–52.

29. Popov VN, Henrard L, Lambin P. Resonant Raman spectra of graphene with point defects. Carbon N Y [Internet]. 2009;47(10):2448–55. Available from: http://dx.doi.org/10.1016/j.carbon.2009.04.043

30. Díez-Betriu X, Álvarez-García S, Botas C, Álvarez P, Sánchez-Marcos J, Prieto C, et al. Raman spectroscopy for the study of reduction mechanisms and optimization of conductivity in graphene oxide thin films. J Mater Chem C. 2013;1(41):6905–12.

31. Ren X, Li J, Chen C, Gao Y, Chen D, Su M, et al. Graphene analogues in aquatic environments and porous media: Dispersion, aggregation, deposition and transformation. Environ Sci Nano. 2018;5(6):1298–340.

32. López-Honorato E, Meadows PJ, Xiao P, Marsh G, Abram TJ. Structure and mechanical properties of pyrolytic carbon produced by fluidized bed chemical vapor deposition. Nucl Eng Des. 2008;238(11):3121–8.

33. Hatakeyama K, Awaya K, Koinuma M, Shimizu Y, Hakuta Y, Matsumoto Y. Production of water-dispersible reduced graphene oxide without stabilizers using liquid-phase photoreduction. Soft Matter. 2017;13(45):8353–6.

34. Cortés-Arriagada D, Toro-Labbé A. Improving As(III) adsorption on graphene based

surfaces: Impact of chemical doping. Phys Chem Chem Phys. 2015;17(18):12056–64.

35. Nyangiwe NN, Khenfouch M, Thema FT, Nukwa K, Kotsedi L, Maaza M. Free-Green Synthesis and Dynamics of Reduced Graphene Sheets via Sun Light Irradiation. Graphene. 2015;04(03):54–61.

36. KUMAR P, SUBRAHMANYAM KS, RAO CNR. Graphene Produced By Radiation-Induced Reduction of Graphene Oxide. Int J Nanosci [Internet]. 2011;10(4–5):559–66. Available from: http://www.worldscientific.com/doi/abs/10.1142/S0219581X11008824

37. Hu X, Zhou M, Zhou Q. Ambient water and visible-light irradiation drive changes in graphene morphology, structure, surface chemistry, aggregation, and toxicity. Environ Sci Technol. 2015;49(6):3410–8.

38. Gengler RYN, Badali DS, Zhang D, Dimos K, Spyrou K, Gournis D, et al. Revealing the ultrafast process behind the photoreduction of graphene oxide. Nat Commun. 2013;4(May):1–5.

39. Li R, Guiney LM, Chang CH, Mansukhani ND, Ji Z, Wang X, et al. Surface Oxidation of Graphene Oxide Determines Membrane Damage, Lipid Peroxidation, and Cytotoxicity in Macrophages in a Pulmonary Toxicity Model. ACS Nano. 2018;12(2):1390–402.

40. Akhavan O, Ghaderi E. Toxicity of graphene and graphene oxide nanowalls against bacteria. ACS Nano. 2010;4(10):5731–6.